\documentstyle[11pt,paspconf,psfig]{article}

\def\etal{et al. }
\begin{document}

\title{Age Estimates for Galaxies in Groups}
\author{Duncan A. Forbes and Alejandro I. Terlevich}
\affil{School of Physics and Astronomy, University of Birmingham,
Birmingham B15 2TT, UK}

\begin{abstract}

We discuss recent developments on the age and metallicity distribution for
early type galaxies in different environments. 

\end{abstract}


\section{Introduction}

A galaxy's environment plays a key role in determining its evolution. For
elliptical galaxies, it is generally 
thought that mergers of disk galaxies in the field
and in groups are the dominant formation mechanism. Elliptical--rich groups
that fall in along filaments create the elliptical--rich clusters we seen
today. 
Tracking the assembly of elliptical galaxies and the evolutionary status of
groups would provide further insight into these processes. 

Until recently
it was very difficult to directly age--date the stars in old stellar
populations due to the age--metallicity degeneracy. 
This degeneracy has now been broken by 
new spectroscopic observations and models 
(e.g. Worthey 1994; Trager \etal 1999). Thus it is now possible to form
an evolutionary sequence of elliptical galaxy formation and to age--date
the ellipticals in different environments.

\section{Deviations from Galaxy Scaling Relations}

In two recent papers (Forbes \etal 1998; Forbes \& Ponman 1999) we showed
that a galaxy's position relative to the fundamental plane and other
scaling relations depends on a galaxy's age. Here age is the central
luminosity weighted age of the galaxy from stellar spectroscopy.
We found
that young ellipticals were brighter with a higher surface brightness. 
Ellipticals that were $\sim$ 10 Gyr old would lie on the FP. 
From simple starburst models, we showed that 
fading central starburst could explain the overall trend. 
The situation was similar for the deviations from 
2D scaling relations such as B--V vs
M$_B$ and Mg$_2$--$\sigma$. Younger galaxies would redden and their Mg$_2$ 
line strengths weaken as the central starburst faded. We concluded that 
these scaling
relations are metallicity--mass sequences with deviations caused by a
galaxy's age.

\section{The Age and Metallicity Distribution of Galaxies}

\begin{figure}
\psfig{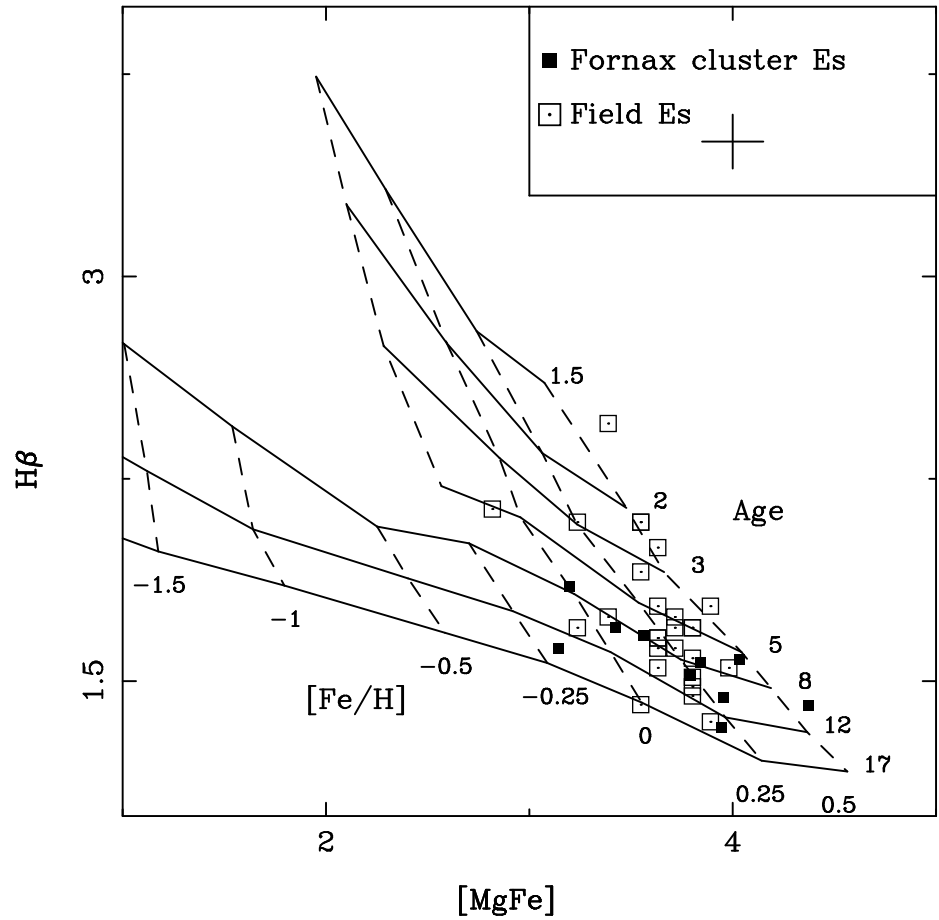}
\caption{
Field and Fornax cluster ellipticals from Gonzalez (1993) and 
Kuntschner \& Davies (1998). The cluster ellipticals are all about 8 Gyrs
old, whereas the field ellipticals appear to scatter in an age sequence at
constant metallicity. 
} \label{fig-1}
\end{figure}

\begin{figure}
\psfig{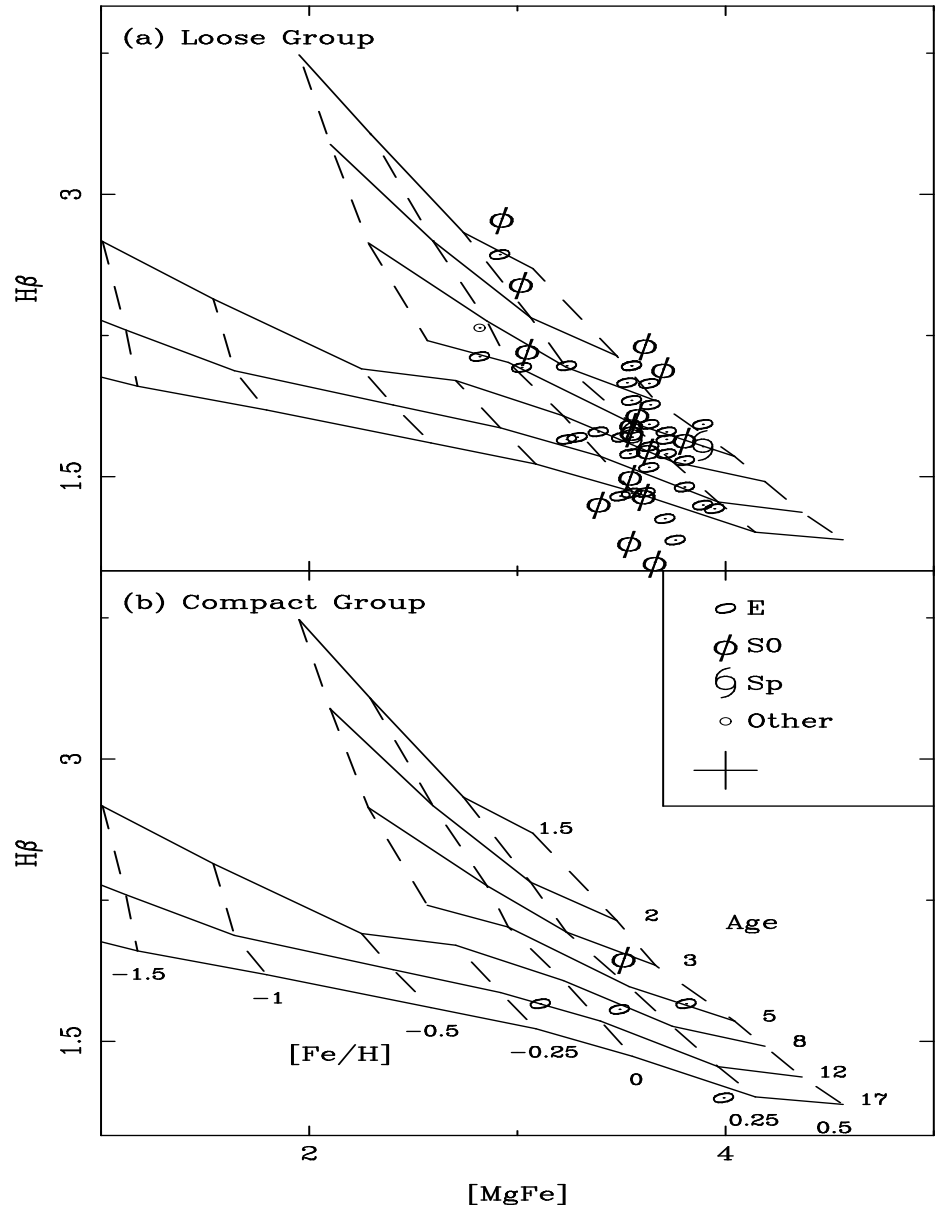}
\caption{
Loose and compact group early type galaxies from the literature. Most loose
group ellipticals appear to be old. There are too few compact group
ellipticals studied to date to describe their age and/or metallicity
distibution. 
} \label{fig-2}
\end{figure}

Perhaps the best, high quality study of field ellipticals is that of
Gonzalez (1993). He obtained new absoprtion line indices for about 40 early
type galaxies in the field, and claimed that when plotted on a 
Worthey (1994) grid of
H$\beta$ vs [MgFe] they generally scatter across a range in ages with
metallicities concentrated around solar.
Another high quality study is that of Kuntschner \& Davies (1998) who
studied early type galaxies in the Fornax cluster. They found all
ellipticals to have a similar age of $\sim$ 8 Gyrs, covering a range in
metallicity. Only the S0 galaxies scattered to young ages in the Worthey
grid. There is certainly support from the Coma cluster that the
colour--magnitude relation is largely a metallicity--mass sequence with the
small scatter due to age effects (Terlevich \etal 1999). 
These field and cluster samples are shown in Fig. 1. Although the cluster
galaxy trends are fairly convincing, more field data is needed to confirm
the Gonzalez claims. 

If field ellipticals appear to describe a sequence in age, while cluster
ellipticals describe a sequence in metallicity (at constant age), how do group
ellipticals behave ?

\noindent
$\bullet$ If group ellipticals resemble cluster ellipticals, then it 
suggests that
`evolutionary' processes have already occured, and must be related to
non--cluster environments, e.g. merging.

\noindent
$\bullet$ If group ellipticals resemble field ellipticals, then it 
suggests that
`evolutionary' processes have yet to occur, and must be related to
cluster environments, e.g. ram pressure stripping, harassment.

An H$\beta$ vs [MgFe] plot for loose groups and compact groups is shown in
Fig. 2. In both cases, most early type galaxies are old ($\sim$
10 Gyr), with some of young age but there are too few to make conclusive
statements. However building up large samples of group galaxies with age
estimates should provide unique clues to their star formation histories, and
in the case of compact groups -- the evolutionary status of the group
itself. 

\acknowledgments

We would like to thank R. Brown, T. Ponman for their
contributions to this work.

\end{document}